\documentclass{elsart}

\usepackage{natbib}
\usepackage{graphicx}
\usepackage{amssymb}
\begin{document}

\begin{frontmatter}

\title{Anisotropic ac dissipation at the surface of mesoscopic superconductors}

\author[label1]{Alexander D. Hern\'{a}ndez}
\author[label1]{Arturo L\'opez-D\'avalos}
\author[label1]{Daniel Dom\'{\i}nguez}

\address[label1]{Centro At{\'{o}}mico Bariloche and Instituto Balseiro,
8400 San Carlos de Bariloche, R\'{\i}o Negro, Argentina.}

\begin{abstract}

In this work we study the ac dissipation of mesoscopic superconductors at microwave frequencies using the time dependent Ginzburg-Landau equations. Our numerical simulations show that the ac dissipation is strongly dependent on the orientation of the ac magnetic field ($h_{ac}$) relative to the dc magnetic field ($H_{dc}$). When $h_{ac}$ is parallel to $H_{dc}$ we observe that each vortex penetration event produces a significant supression of the ac losses because the imaginary part of the ac susceptibility as a function of $H_{dc}$ increases before the penetration of vortices, and then it decreases abruptly after vortices have entered into the sample. In the second case, when $h_{ac}$ is perpendicular to $H_{dc}$, we observe that the jumps in dissipation occur at the same values of $H_{dc}$ but are much smaller than in the parallel configuration. The behavior of the dissipation in the perpendicular configuration is similar to previous results obtained in recent microwave experiments using mesoscopic litographed squares of Pb \cite{Her_JLTP04}.

\end{abstract}


\end{frontmatter}

In macroscopic samples the ac dissipation is generated by the motion of vortices that are typically located some micrometers away from the sample surface, while the dissipation due to surface currents itself can be neglected \cite{Clem_PRL91}. The opposite scenario appears in  mesoscopic samples where the  contribution of the surface currents dominates the dissipation, and vortices play a secondary role due to its confinement by the surface currents \cite{Her_PRB02b}.
Recently, the ac dissipation of an array of mesoscopic square superconductors has been  measured at microwave frequencies \cite{Her_JLTP04}. The resulting ac dissipation presents a non monotonous and oscillating dependence with magnetic field that could be associated with the modulation of the vortex entrance events on the surface  dissipation.

\begin{figure}[htb] 
\begin{center}
\includegraphics[width=0.7\linewidth]{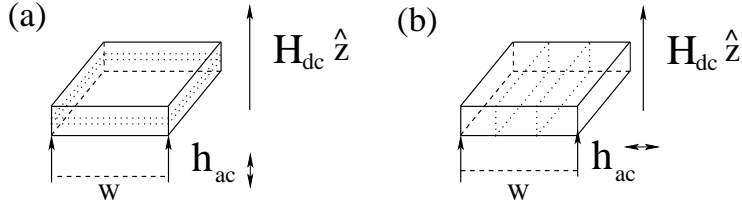}
\caption{The studied magnetic field configurations, (a) $h_{ac} \parallel H_{dc}$ and (b) $h_{ac} \perp H_{dc}$. Doted lines show planes of circulation of the induced ac currents.}
\label{geometry} 
\end{center}
\end{figure}

In this paper we study the importance of the relative orientation of the ac and dc magnetic fields ($h_{ac}$ and $H_{dc}$) on the surface dissipation of mesoscopic superconductors.
 Fig. \ref{geometry}(a) represents the parallel configuration, $h_{ac}\parallel H_{dc}$, and Fig. \ref{geometry}(b) shows the perpendicular configuration, $h_{ac}\perp H_{dc}$. The doted lines show planes,  perpendicular to $h_{ac}$, where the induced ac currents circulate. 
In the present study we neglect the influence of the demagnetization field assuming that the exact treatment thereof only leads to quantitative corrections, while the  qualitative behavior remains the same.

To study the effects of the anisotropy on the ac response we made numerical simulations of the time-dependent Ginzburg-Landau equations in three-dimensions. The normalized equations are, see Refs. \cite{Her_PRB02b} and \cite{Gropp_JCP96},
{\setlength\arraycolsep{1pt}
\begin{eqnarray}
\frac{\partial \Psi}{\partial t}&=&-\frac{1}{12}\biggl[\Bigl(-\imath\nabla-\vec{A}\Bigr)^2\Psi
+\Bigl(1-T\Bigr)\Bigl(|\Psi|^2-1 \Bigr)\Psi \biggr]  \nonumber \\
\frac{\partial \vec{A}}{\partial t}&=&(1-T)\Re e \Bigl[
\Psi^*(-\imath\nabla-\vec{A})\Psi \Bigr]-\kappa^2(\nabla \times \vec{B}), \nonumber
\end{eqnarray}}
where $\Psi$ is the order parameter, $\vec A$ is the vector potential and T is the temperature. In the above equations, lengths have been scaled in units of the coherence length $\xi(0)$, times in units of $t_0=4\pi \sigma_n \lambda_L^2/c^2$,
$\vec{A}$ in units of $H_{c2}(0) \xi (0)$, $\Psi$ in units of
$\Psi_{\infty}(T)=[mc^2/8\pi e^2\lambda(T)^2]^{1/2}$ and temperature in units
of $T_c$.

The boundary conditions for this problem are
\begin{eqnarray} 
(\Pi \Psi)^\perp &=&(\nabla -i\vec{A})^\perp \Psi =0 \nonumber \\
\vec{\nabla} \times \vec{A}(t)&=&H_{dc}\hat{z} + [h_{ac}\cos(\omega t)]\hat{\mu},
\nonumber
\end{eqnarray}
where $H_{dc}$ is oriented in the $\hat{z}$ direction (see Fig. \ref{geometry}) and $\hat{\mu}$ is the direction of the ac magnetic field. When $\hat{\mu}=\hat{z}$ we obtain the parallel configuration of Fig. \ref{geometry}(a) and when $\hat{\mu}=\hat{x}$ or $\hat{\mu}=\hat{y}$ we obtain the perpendicular configuration shown in Fig. \ref{geometry}(b). 
In this work we study the behavior of the imaginary part of the ac magnetic susceptibility $\chi "(H_{dc},\omega ) = \frac{1}{\pi h_{ac}} \int_0^{2 \pi} M(t)\sin(\omega t)$. The numerical method to solve the three-dimensional set of differential equations is explained elsewhere, see Refs. \cite{Her_PRB02b} and \cite{Gropp_JCP96}.

\begin{figure}[htb] 
\begin{center}
\includegraphics[width=0.60\linewidth]{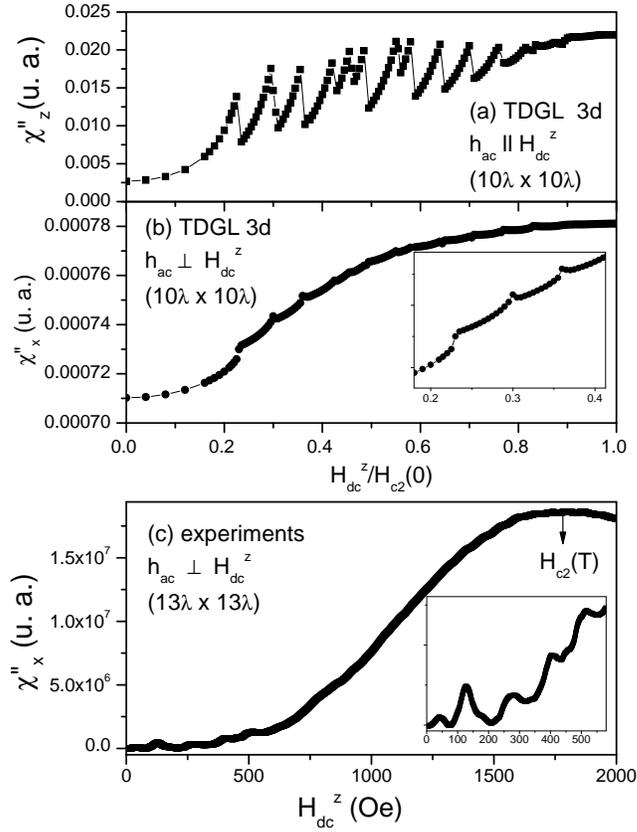}
\caption{(a) and (b) are numerical results for the two relative orientations between $h_{ac}$ and $H_{dc}$. In (c) we show recent experimental results obtained in mesoscopic samples (taken from Ref. \cite{Her_JLTP04}).}
\label{tdgl-3d} 
\end{center}
\end{figure}

In Figs. \ref{tdgl-3d}(a) and (b) we show numerical results obtained for samples of size $10\lambda \times 10 \lambda \times 1\lambda$, where $\lambda$ is the zero temperature penetration depth. We considered here $\kappa=2$ and T=0.5. 
Fig. \ref{tdgl-3d}(a) shows the imaginary part of the ac susceptibility ($\chi"$) as a function of $H_{dc}$ for the parallel configuration ($h_{ac} \parallel H_{dc}$). We observe that the dissipation decreases after each vortex penetration event, and increases continuously when the number of vortices is constant. Vortices enter following the square symmetry of the sample in packages of four vortices \cite{Her_PRB05}. However, thermal  fluctuations and surface imperfections can allow the number of vortices to increase one by one as was shown in Ref. \cite{Her_PRB05}. The results of Fig. \ref{tdgl-3d}(a) are valid for perfect samples at zero temperature. A similar qualitative  behavior has been observed in Ref. \cite{Her_PRB02b} where a two-dimensional approach was used.

Fig. \ref{tdgl-3d}(b) shows results obtained using the perpendicular configuration ($h_{ac} \perp H_{dc}$). We observe that the vortex penetration events occur at the same magnetic fields as in Fig. \ref{tdgl-3d}(a). This is a consequence of the small value of $h_{ac}$ compared with $H_{dc}$ ($h_{ac}=5\times 10^{-4}H_{c2}\ll H_{dc}$). In the linear regime, the magnitude of the dc magnetic field defines in both cases the condition for vortex penetration. Apart from this similarity, the general behavior of $\chi "$ vs. $H_{dc}$ in Fig. \ref{tdgl-3d}(b) is different to the one in Fig. \ref{tdgl-3d}(a). In the perpendicular configuration (Fig. \ref{tdgl-3d}(b)) we see that vortex entrances  produce smaller jumps in $\chi"(H_{dc})$ compared with the jumps occurring in the parallel configuration (Fig. \ref{tdgl-3d}(a)). In Fig. \ref{tdgl-3d}(b) we only observe a small decrease in $\chi "$ and a change of slope at $H_{p,i}$ (the $i$-th penetration field).

Fig. \ref{tdgl-3d}(c) allows a direct comparison between numerical and recent experimental results. In Ref. \cite{Her_JLTP04} the ac dissipation of an array of mesoscopic square superconductors was measured at microwave frequencies using the perpendicular configuration.  A simple qualitative comparison shows that the experimental results are better described by the numerical results of Fig. \ref{tdgl-3d}(b) which corresponds to the perpendicular configuration.
Spectra of Ref. \cite{Her_JLTP04} were acquired at 9.5GHz while the numerical simulations corresponds to a frequency of $\omega=0.09\upsilon_o$, where $\upsilon_o=1/t_o\sim 100-1000$GHz in dirty superconductors.

The reason for the appearance of the anisotropy is clear from Figs. \ref{geometry}(a) and \ref{geometry}(b). In both configurations, above $H_{p,1}$, vortices are oriented in the direction of the dc magnetic field. The ac magnetic field induce currents, responsible of the ac dissipation, circulate in planes perpendicular to $h_{ac}$. In the parallel configuration (Fig. \ref{geometry}(a)) the ac currents circulates on the faces of the sample where the nascent vortices \cite{Wal_PL73} are located. These ac currents generate periodic compresions and expansions of the vortices that are localized inside the sample, but the dissipation is mainly due to the nascent  vortices. On the other hand, in the perpendicular configuration (Fig. \ref{geometry}(b)), the ac currents circulate  mainly on the top and bottom faces of the sample where the order parameter is depleted due to the vortex cores and not due to nascent vortices. The dissipation is larger  than in the previous case and has a different dependence with $H_{dc}$. 
In conclusion, the orientation of the vortices along the $H_{dc}$ direction and the change in orientation of the induced ac currents in both configurations generate the anisotropy of the response.

We acknowledge support from CNEA, CONICET (PIP2005-5596), and ANPCyT (PICT 2003-13829 and PICT 2003-13511).

\end{document}